\documentclass[aps,prl,preprint,groupedaddress,superscriptaddress, 
nobibnotes,nofootinbib,showkeys]{revtex4-1}
\usepackage{float}
\usepackage{graphicx}
\usepackage{graphics}
\usepackage{epsfig}
\usepackage{latexsym}
\usepackage{colordvi}
\usepackage{amsmath}
\usepackage{amssymb}
\usepackage{slashed}
\usepackage{xcolor}
\usepackage[title]{appendix}

\newcommand{\be}{\begin{equation}}
\newcommand{\ee}{\end{equation}}
\newcommand{\bea}{\begin{eqnarray}}
\newcommand{\eea}{\end{eqnarray}}

\def\IB{\relax\hbox{$\inbar\kern-.3em{\rm B}$}}
\def\IC{\relax\hbox{$\inbar\kern-.3em{\rm C}$}}
\def\ID{\relax\hbox{$\inbar\kern-.3em{\rm D}$}}
\def\IE{\relax\hbox{$\inbar\kern-.3em{\rm E}$}}
\def\IF{\relax\hbox{$\inbar\kern-.3em{\rm F}$}}
\def\IG{\relax\hbox{$\inbar\kern-.3em{\rm G}$}}
\def\IGa{\relax\hbox{${\rm I}\kern-.18em\Gamma$}}
\def\IH{\relax{\rm I\kern-.18em H}}
\def\IK{\relax{\rm I\kern-.18em K}}
\def\IL{\relax{\rm I\kern-.18em L}}
\def\IP{\relax{\rm I\kern-.18em P}}
\def\IR{\relax{\rm I\kern-.18em R}}
\def\IZ{\relax{\rm Z\kern-.5em Z}}

\begin{document}

\preprint{\parbox[b]{1in}{ \hbox{\tt PNUTP-24/A01}  }}
\title{Magnetic axion vortex and very light QCD axions}

\author{Deog Ki Hong}
\email[]{dkhong@pusan.ac.kr} 
\affiliation{Department of
Physics,   Pusan National University,
             Busan 46241, Korea}
             
\affiliation{Extreme Physics Institute, Pusan National University, Busan 46241, Korea}
\author{Stephen J. Lonsdale}
\email[]{stephenjlonsdale@gmail.com} \affiliation{Department of
Physics,   Pusan National University,
             Busan 46241, Korea}

\affiliation{Extreme Physics Institute, Pusan National University, Busan 46241, Korea}
\date{\today}

\begin{abstract}
We show that a stable vortex soliton, carrying a constant magnetic flux, exists in the homogeneous medium of axions with a constant time-derivative. Axions can be bound in the vortex, having energy less than the axion mass. If the observed magnetic fields in galaxies are those of the vortex, the axion-photon coupling has to be smaller than $10^{-17}\,{\rm GeV^{-1}}$. Otherwise, axions decay too quickly to constitute dark matter in galaxies. 
\end{abstract}


\maketitle



\section{Introduction}
Among various candidates for dark matter in our universe, the quantum chromodynamics (QCD) axion is still the most compelling one, because it not only is consistent with the stringent analysis of the cosmic microwave background~\cite{Planck:2018nkj} but also elegantly solves the strong CP problem of QCD~\cite{Peccei:1977hh}. 
 A great effort is currently being made worldwide to search for axion dark matter, whose sensitivity is now getting close to theoretically interesting levels~\cite{Semertzidis:2021rxs}. 
The salient property of axion is that it couples in general to photons through the Adler-Bell-Jackiw (ABJ) anomaly of $U(1)_{\rm PQ}$ symmetry, introduced to solve the strong CP problem. Such anomalous axion-photon coupling provides various means to detect it, even though axion is electrically neutral and has no spin~\cite{Sikivie:2020zpn}.  

Axion electrodynamics that describes the axion-photon interaction has been studied in depth in the literatures. One of the pertinent features of axion electrodynamics is that photons propagating in the background of axion fields are unstable for wavelengths longer than a critical value, $\lambda_c\sim (g_{a\gamma}\dot a)^{-1}$, inversely proportional to the product of the axion-photon coupling, $g_{a\gamma}$, and the time derivative of the background axion fields, $\dot a$~\cite{Finelli:2000sh}. In this study we show that the axion electrodynamics admits a global vortex-like soliton of size $\lambda_c$, supported by magnetic flux, in a medium of axions or axion-like particles (ALP) with constant a time-derivative. Axions having a constant time-derivative to carry a large kinetic energy, known as axion kination, might exist in the early universe by the kinetic misalignment~\cite{Co:2019jts}. In this case the background axion field probes the full range of the potential and axion quanta will produce copiously near the maximum of the axion potential, known as axion fragmentation~\cite{Fonseca:2019ypl}. Here we consider instead the axion field, trapped to oscillate or moving near the minimum of the potential, and focus on the moment $\Delta t\ll m_a^{-1}$ during which the acceleration of the background axion field is negligible, $\ddot a\approx0$, so that the axion velocity is almost constant to support the vortex soliton in the background magnetic field.

The vortex solution is shown to be stable against small fluctuations. The axion normal modes inside the vortex have frequencies smaller than the axion mass. They are therefore confined in the vortex.  The axion quanta, confined inside the vortex, move along the vortex, carrying a constant momentum as well as an angular momentum along the vortex, which turn out to be extremely stable and might explain the observed angular momentum of galaxies.

\section{Axion electrodynamics}
When the ABJ anomaly is present, axions or axion-like particles couple to photons with a strength, $g_{a\gamma}$, proportional to the anomaly. The Lagrangian for the photon fields coupled to axions becomes 
\begin{equation}
	{\cal L}=-\frac14F_{\mu\nu}F^{\mu\nu}-\frac{g_{a\gamma}}{8}a\,\epsilon^{\mu\nu\rho\sigma}F_{\mu\nu}F_{\rho\sigma}\,-ej^{\mu}A_{\mu}+{\cal L}_{a}\,,
\end{equation}
where the field strength tensor $F_{\mu\nu}=\partial_{\mu}A_{\nu}-\partial_{\nu}A_{\mu}$\, and the axion Lagrangian density is given as 
\begin{equation}
{\cal L}_a=\frac12\partial_{\mu}a\,\partial^{\mu}a-V(a)\,,	
\end{equation}
where $V(a)$ is the axion potential due to the explicit breaking of the $U(1)_{\rm PQ}$\, such as the quark mass~\cite{GrillidiCortona:2015jxo}.
The Maxwell equation becomes now~\cite{Sikivie:2020zpn}  
\begin{equation}
\partial_{\mu}F^{\nu}+\frac12g_{a\gamma}\epsilon^{\mu\nu\rho\sigma}\partial_{\mu}\left(aF_{\rho\sigma}\right)=ej^{\nu}\,,	
\end{equation}
supplemented with the usual Bianchi identity
\begin{equation}
\epsilon^{\mu\nu\rho\sigma}\partial_{\nu}F_{\rho\sigma}=0\,.	
\end{equation}
In terms of the electric and magnetic fields, $E^i=F^{i0}$ and $B^i=\frac12\epsilon^{ijk}F_{jk}$\,, we have 
\begin{eqnarray}
\vec\nabla\cdot\left(\vec E-g_{a\gamma}a\vec B\right)=ej^0\,,\quad 
\vec\nabla\times\left(\vec B+g_{a\gamma}a\vec E\right)-\frac{\partial}{\partial t}\left(\vec E-g_{a\gamma}a\vec B\right)=e\vec j\;	\label{modified}\\
\vec\nabla\cdot\vec B=0\,,\quad \vec\nabla\times\vec E+\frac{\partial}{\partial t}\vec B=0\,.
\label{bianchi}
\end{eqnarray}

\section{Vortex solution}
\begin{figure}[ht]
\center \includegraphics[scale=0.5]{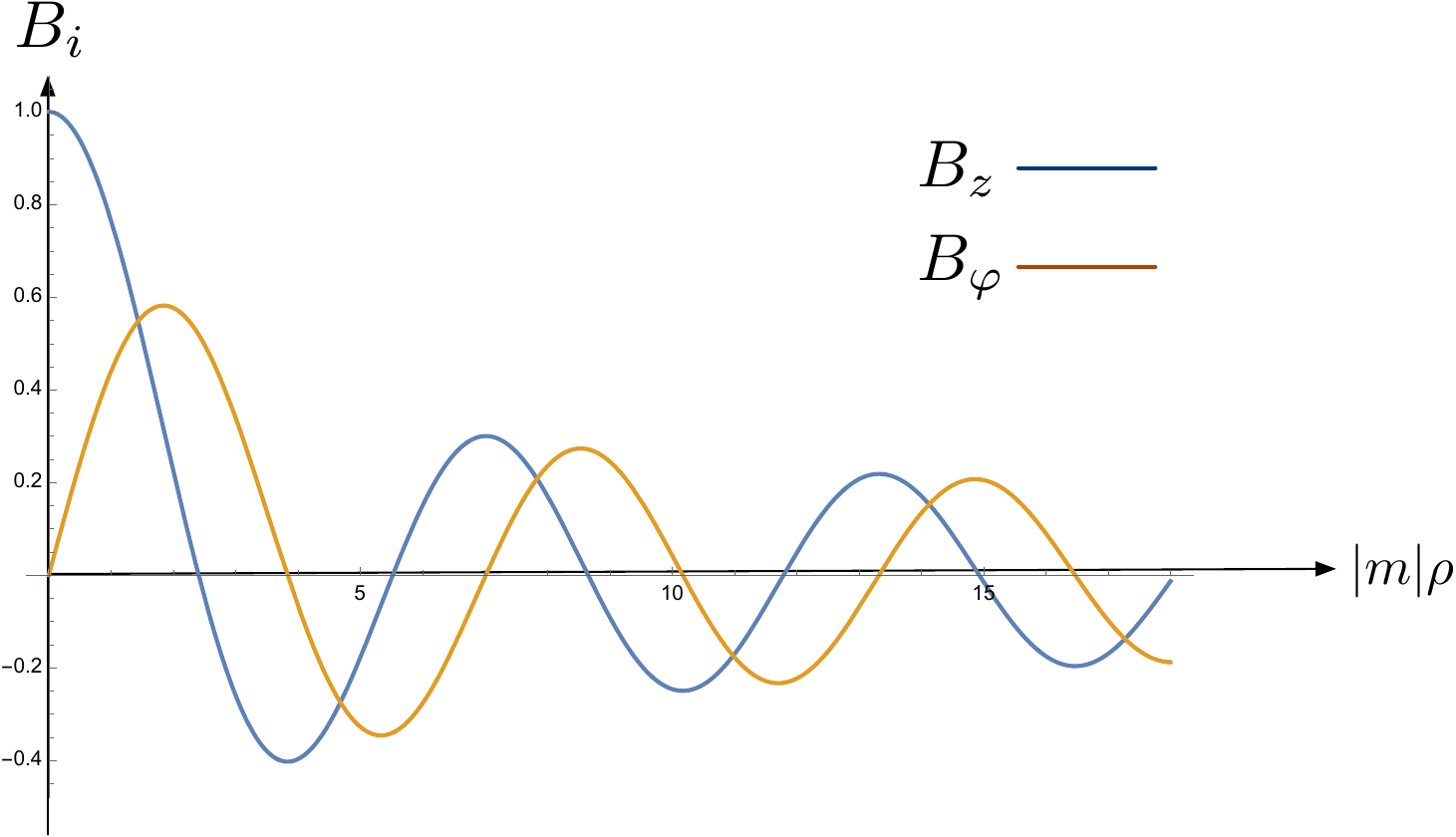}
\caption{The components of the vortex magnetic field, $B_i$ ($i=z,\varphi$) for a given magnetic flux $\Phi$, in units of $\Phi\, m^2/{\cal N}$ with $m=g_{a\gamma}\dot a<0$.}
\label{vortex}
\end{figure}

In the presence of the electromagnetic fields, the equation of motion for axions becomes 
\begin{equation}
\partial^2a+V'(a)=-g_{a\gamma}	\vec E\cdot\vec B\,,
\end{equation}
where non-vanishing $\vec E\cdot\vec B$ sources axions.
When there is no source, a homogeneous cold medium of axions like cosmic dark matter, $\vec\nabla a\approx 0$, behaves as a collective mode,  
\begin{equation}
a(t)\approx A_0\sin\left(m_at\right)\,,	
\end{equation}
where we approximate $V^{\prime}(a)\approx m_a^2a$\,, keeping only the axion mass term. 
Now, in the absence of charged fields,  $j^{\mu}=0$, the modified Maxwell equation, Eq.~(\ref{modified}), becomes for the axion collective mode 
\begin{equation}
\vec\nabla\cdot\vec E=0\,,\quad\vec\nabla\times\vec B-\frac{\partial}{\partial t}\vec E=
-g_{a\gamma}\dot a\vec B\,,	
\label{maxwell}
\end{equation}
where $\dot a$ denotes the time derivative of the axion field. 
For axions near the minimum of the potential or for $|t|\ll m_a^{-1}$, the axion velocity $\dot a$ is almost constant.
As we will show, when $\dot a$ is a constant~\cite{Hong:2022nss}, the Maxwell equations, Eq.'s~(\ref{bianchi}) and (\ref{maxwell}) admit  a soliton solution, supported by a finite magnetic flux that is topologically conserved. We therefore look for a static solution of minimum energy for a given magnetic flux, $\Phi$, along the $z$ direction. Since the electric field does not contribute to the magnetic flux but costs energy, our Ansatz in the cylindrical coordinates will be
\begin{equation}
\vec E=0\,,\quad \vec B=\left(0, B_{\varphi}(\rho), B_z(\rho)\right)\,,	
\end{equation}
where $\rho$ is the distance from the $z$ axis where the soliton is located.
In axion electrodynamics, the magnetic field sources itself, producing a current $\vec J=-m\vec B$ with $m=g_{a\gamma}\dot a$ even in the absence of charged particles. We ask then how the magnetic field along the vortex should be distributed to minimize its energy,  
\begin{equation}
{\cal E}=\!\int{\rm d}^3x\left\{\frac12{\vec B}^2-\vec A\cdot\vec J\right\}
=\int{\rm d}^3x\left\{\frac12\left(\vec B+m\vec A\right)^2-\frac12m^2{\vec A}^2\right\}\ge \int\!{\rm d}^3x\frac12m\,{\vec A}\cdot\vec B\,,
\label{energy}
\end{equation}
where $\vec A$ is the vector potential and we have used $\vec J=-m\vec B$\, in the second equality.  We see that the energy is saturated by configurations that satisfy $\vec B=-m\vec A$ in the Coulomb gauge, $\vec \nabla\cdot \vec A=0$, and is proportional to the magnetic helicity of the field~\cite{Cornwall:1997ms}. They satisfy the Maxwell equation (\ref{maxwell}) consequently. The minimum energy configuration should then satisfy 
\begin{equation}
B_{\varphi}^{\prime\prime}+\frac{1}{\rho}B_{\varphi}^{\prime}-\left(\frac{1}{\rho^2}-m^2\right)B_{\varphi}=0\,; \quad B_{z}^{\prime\prime}+\frac{1}{\rho}B_{z}^{\prime}+m^2B_{z}=0
\end{equation}
where the prime denotes the derivative with respect to the radial coordinate, $\rho$.
We find the minimum energy configuration for a given finite magnetic flux $\Phi$ to be 
\begin{equation}
B_{\varphi}(\rho)=-m|m|\,\frac{\Phi}{{\cal N}} J_1\left(\left|m\right|\rho\right),	\,\,
B_z(\rho)=m^2\,\frac{\Phi}{{\cal N}} J_0(|m|\rho)\,.
\label{magnet}
\end{equation}
where $J_n(x)$'s are the $n$-th order Bessel functions and ${\cal N}=2\pi\int_0^{x_c}xJ_0(x){\rm d}x$ for a given radial cutoff $\rho_c=x_c/|m|$ that regulates the infrared divergences with $x_c\gg1$\,. We call the solution a magnetic axion vortex because it is cylindrically symmetric and carries a magnetic flux, spread around its center. The magnetic fields are interweaved along the vortex. 
 (See Fig.\,\ref{vortex}.) The magnetic axion vortex is topologically stable, supported by both magnetic flux and the homogeneous axion field with constant time-derivative. 
We note that the tension or the energy of magnetic vortex per unit length is logarithmically divergent when its radial size is infinite, which must be therefore cutoff at $\rho_c$, either the size of the horizon or the scale where $\dot a$ ceases to be constant. 

\section{Back reactions to axions}
Once the vortex is formed, it is subject to fluctuations, which may destabilize it.  To see the fate of the soliton when the background axion fluctuates\,\footnote{The electromagnetic fields will fluctuate as well, but they would not destabilize the vortex since the Maxwell Lagrangian is quadratic in photon fields.}, we consider a small fluctuation of axions around the homogeneous background with constant velocity, $\dot a_0$: 
\begin{equation}
	a={a}_0+\delta a\,.
\end{equation}
 Then, $\dot a$ is no longer constant but changes in time and the (time-dependent) magnetic fields of vortex will create electric fields by the Faraday's law in Eq.~(\ref{bianchi}). 
Treating the fluctuations, $\delta a$, small, we solve the Faraday equation perturbatively to find at the leading order in $\delta a$, with $m=g_{a\gamma}{\dot a_0}$\,,
\begin{equation}
E_z(\rho)=g_{a\gamma}\,{\delta \ddot a}\,\frac{\Phi}{\cal N}\left[J_0(|m|\rho)-|m|\rho \,J_1(|m|\rho)\right],\quad
E_{\varphi}(\rho)=-g_{a\gamma}\,{\delta \ddot a}\,\frac{\Phi}{\cal N}\,m\rho\, J_0(|m|\rho)\,.
\end{equation}
In the leading order of $\delta a$ the magnetic vortex carries a non-vanishing axion source~\footnote{Since both the axion source, $\vec E\cdot\vec B$, and axions are parity odd, the source term, Eq.\,(\ref{source}), should be independent of the sign of $m$ or $\dot a_0$ in the leading order in $\delta a$.}
\begin{equation}
\vec E\cdot\vec B=g_{a\gamma}\,{\delta \ddot a}\,m^2\left(\frac{\Phi}{\cal N}J_0(|m|\rho)\right)^2\,,
\label{source}
\end{equation}
which renormalizes the axion kinetic term. 
The small fluctuation of axions hence satisfies~\footnote{The gradient of the axion field will contribute to the axion source term but only at the higher order. When the axion source term, $\vec E\cdot\vec B$, is absent, this equation is precisely the Mathieu equation that leads to the axion fragmentation, studied in~\cite{Fonseca:2019ypl}\,.
} \begin{equation}
H(|m|\rho)\,\delta\ddot a-\nabla^2\,\delta a
+V^{\prime\prime}(a_0)\,\delta a=0\,,
\end{equation}
where 
\begin{equation}
H(|m|\rho)=1+g_{a\gamma}^2\,m^2\left(\frac{\Phi}{\cal N}J_0(|m|\rho)\right)^2\,.	\end{equation}

Since we are considering the fluctuations near the potential minimum, where $|\ddot a|\ll m_a|\dot a|$, we take $V^{\prime\prime}(a_0)\approx m_a^2$.
Now we note that, as $H(|m|\rho)>1$ for all $\rho$ and approaches to $1$ at infinity~\footnote{The fact that $H(|m|\rho)$ should be greater than one is easy to understand. When the axion-photon interaction is turned on, accelerating axions will radiate the electromagnetic waves, which results in increasing the inertia of axions, namely $H>1$.}, there always exist normalizable modes with $\omega<m_a$.  
We try therefore as an Ansatz 
\begin{equation}
\delta a=\theta(t)R(|m|\rho)\, 
\end{equation}
and get, after rescaling $|m|\rho$ to be $\rho$, 
\begin{equation}
	-\nabla^2R(\rho)-\frac{\omega^2}{m^2}H(\rho)R(\rho)=-\lambda R(\rho)\,,
	\label{sch}
\end{equation}
where $\omega^2=-\ddot\theta/\theta$ and $\lambda=m_a^2/m^2$. 
The problem we need to solve is then equivalent to finding  $S$-wave energy eigenstates for a given eigenvalue $-\lambda$ in two-dimensional Schr\"odinger equation under a potential $V(\rho)=-\omega^2/m^2H({\rho})$, while adjusting the depth of the potential  by ${\bar\omega}^2=\omega^2/m^2$  (See Fig.~2). 
\begin{figure}[ht]
\centering 
\includegraphics[scale=0.4]{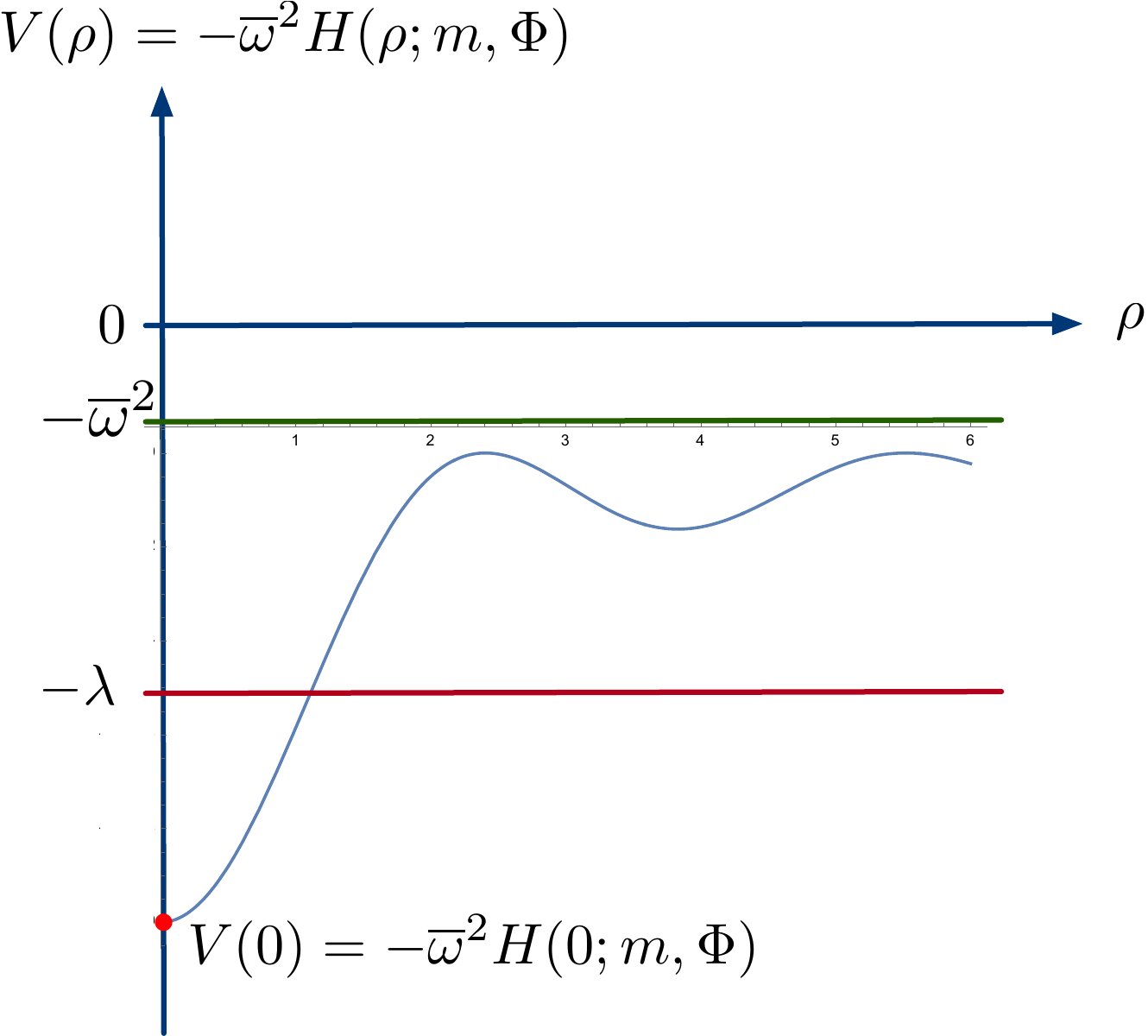}
\caption{The potential energy of two dimensional Schr\"odinger equation whose depth shifts with ${\overline\omega}^2\equiv\omega^2/m^2$.}
\label{shape}
\end{figure}
 We see that, when $\omega>m_a$, namely $-{\bar\omega}^2<-\lambda$, no bound states exist and hence the (excited) magnetic vortex radiates away the high frequency axions to infinity, losing its energy. 
The vibrating vortex can be stable, however, 
for a certain set of discrete frequencies, $\omega_n<m_a$ with $n=0,1,2,\cdots$ in the increasing order, given by the Schr\"odinger equation, Eq.~(\ref{sch}). 
We solve numerically the Schr\"odinger equation to plot the profiles of the a few low-lying $S$-wave bound states  and their energy eigenvalues as a function the magnetic flux $\Phi$ for a given $m$, namely $\lambda=m_a^2/m^2$. (See Fig.~\ref{shape}.)
Since both $H(\rho)$  and $\lambda$ are always positive, $\omega^2<0$ does not admit any bound-state solutions. There is therefore no tachyonic instability in the (perturbative) fluctuations of the magnetic axion vortex~\footnote{The vortex might be unstable non-perturbatively under tunneling to the axion configuration, $a_c$, at which $V^{\prime\prime}(a_c)<\omega^2$\,.}. 
\begin{figure}[ht]
\centering 
\includegraphics[scale=0.32]{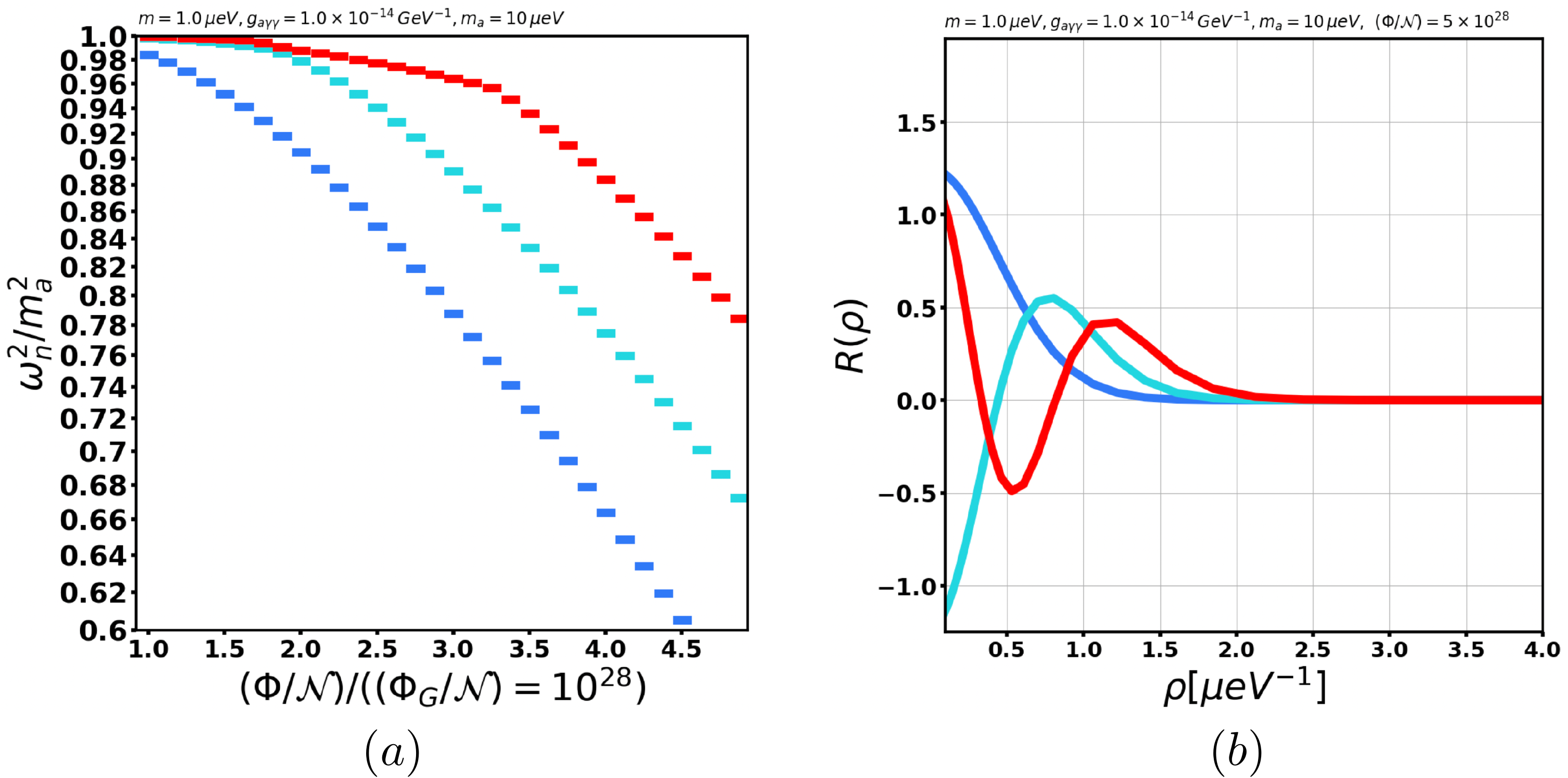}
\caption{For an illustration we take $m=1\mu{\rm eV}$, $g_{a\gamma}=10^{-14}\,{\rm GeV^{-1}}$ and $m_a=10\mu{\rm eV}$ for the flux $10^{28}<\Phi/{\cal N}<5\times 10^{28}$ to show a few low-lying $S$-wave eigenstates:  (a) The eigenfrequencies $\omega_n$ ($n=0,1,2$) of normal modes. (b) The corresponding radial profiles of normal modes.  }
\label{shape}
\end{figure}

\begin{figure}[ht]
\centering 
\includegraphics[scale=0.32]{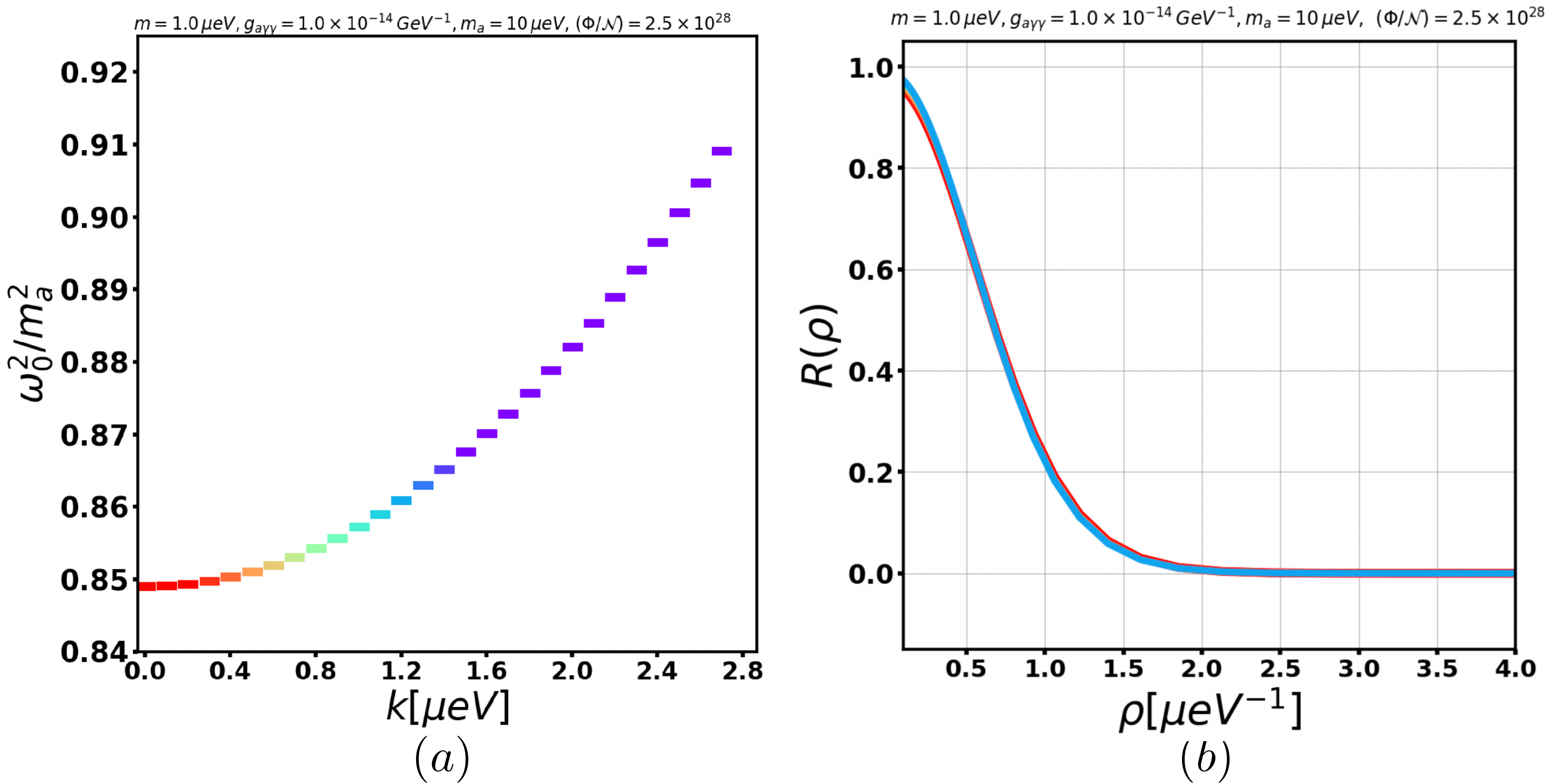}
\caption{(a) The energy dispersion for the ground state axion moving along the vortex 
for $m=1\mu{\rm eV}$, $g_{a\gamma}=10^{-14}\,{\rm GeV^{-1}}$ and $m_a=10\mu{\rm eV}$ with the flux $\Phi/{\cal N}=5\times 10^{28}$. 
(b) The corresponding radial profile with non-vanishing $k_z$}\,.
\label{dispersion}
\end{figure}

As the vortex is invariant under the time translation and the azimuthal rotation, the general energy eigenstates of bound axions take a following form,
\begin{equation}
\delta a_{nk_zl}(t,\rho,\phi,z)=\theta(\omega t -k_zz)e^{il\phi}R_{nk_zl}(\rho)\,,	
\end{equation}
where $\theta$ is a real function to denote a freely traveling wave along the vortex and $\omega$ is the bound state energy, depending on the momentum, $k_z$, along the $z$ axis and the azimuthal angular momentum, $l=0,1,2,\cdots$. 
Since the energy eigenstate for non-zero $k_z$ corresponds to the energy eigenstate at rest, having the axion mass shifted to $m_a^2+k_z^2$ or $\lambda\to \lambda+k_z^2/m^2$ in Eq.~(\ref{sch}), the energy eigenvalue will increase for axions traveling along the vortex to have a nontrivial dispersion relation, $\omega(k_z)$, until $k_z$ reaches its maximum.   
We plot in Fig.~\ref{dispersion} the energy dispersion for the ground state axions, $\omega_{0}(k)$, and its radial profiles for different $k_z$.  

\section{Radiation of photons and axion zero modes}
\begin{figure}[ht]
\vskip 0.2in
\centering 
\includegraphics[scale=0.43]{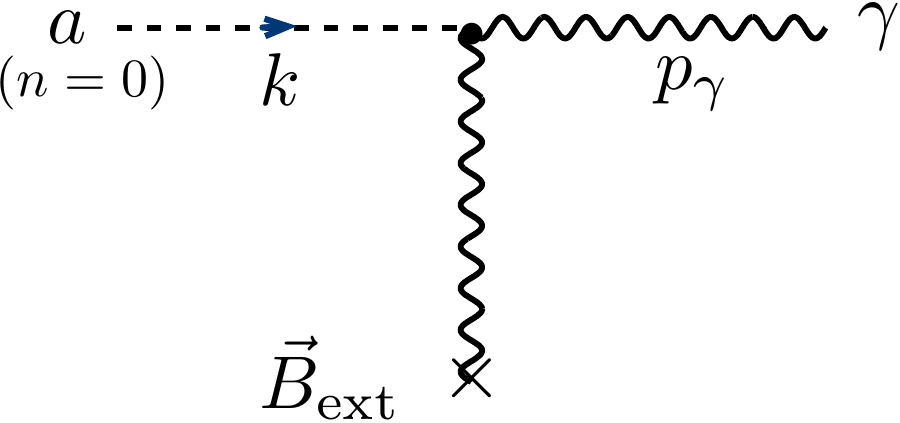}
\caption{The ground state axion decays into a single photon under the magnetic field of the axion vortex.}
\label{axion_decay}
\end{figure}

In the previous section we show that the vortex soliton is stable against the small fluctuations of axions. 
Since the axions do decay into photons, however, the vortex soliton will be unstable against decaying into photons. 
The ground state axion, for example,  will decay into a single photon (See Fig.~\ref{axion_decay})\,\footnote{Axions can decay into a single photon under an inhomogeneous magnetic field, where three-momentum is no longer conserved.}.
The ground state axion is freely traveling along the vortex, while localized radially: 
\begin{equation}
\delta a_{0k_z0}=\sin\left(\omega t-k_z z\right)R_{0k_z0}(\rho)=\!\int\frac{d^2k_{\perp}}{(2\pi)^2}\,
C_{k_z}(\vec k_{\perp})\,e^{i(\omega t-k_zz-\vec k_{\perp}\cdot\vec x_{\perp})}\,,
\end{equation}
where $C_{k_z}(\vec k_{\perp})$ is the Fourier-transform of the radial wave-function of the ground state axion moving along the vortex. 
The transition matrix from the axion of momentum $k^{\mu}=(\omega,\vec k_{\perp},k_z)$ to a single photon under the magnetic field of the axion vortex, Eq.\,(\ref{magnet}), is given at the leading order as, for the $A_0=0$ gauge,  
\begin{eqnarray}
T(a\to\gamma)&=&-ig_{a\gamma}\int \!d^4x\left<p_{\gamma};\vec \epsilon(\vec p_{\gamma})\middle|a(x)\partial_0\vec A(x)\cdot \vec B(\vec x_{\perp})
\middle|\omega, \vec k_{\perp},k_z\right>\,.	\nonumber\\
&=& g_{a\gamma}\omega\int \!d^2x_{\!\perp}\,\vec{\epsilon^*}(\vec p_{\gamma})\cdot \vec B(\vec x_{\perp})\,e^{i(\vec k_{\perp}-\vec p_{\perp})\cdot\vec x_{\perp}}(2\pi)^2\delta^2(p_{\shortparallel}-k_{\shortparallel})\,,
\end{eqnarray}
where $\vec x_{\perp}$ denotes the coordinates perpendicular to the vortex and the subscript $\shortparallel$ denotes the time and $z$ component.
Using the following identities for the $n$-th order Bessel functions~\cite{arfken}, 
\begin{eqnarray}
\int_0^{2\pi}{d}{\varphi}\,e^{ in\varphi+i|\vec q_{\perp}|\rho\cos\varphi}&=& 2\pi i^n J_n\left(|\vec q_{\perp}|\rho\right)\,,	\\
\int_0^{\infty}\rho\, d \rho \,J_n\left(|m|\rho\right)J_n\left(|\vec q_{\perp}|\rho\right)&=&\frac{1}{|m|}\delta\left(|m|-|\vec q_{\perp}|\right)\,,
\end{eqnarray}
one gets
\begin{equation}
T(a\to\gamma)=\frac{1}{\sqrt{2}}g_{a\gamma}\omega \,m|m|\frac{\Phi}{\cal N}	(2\pi)^3\frac{1}{|m|}\delta\left(|m|-|\vec p_{\perp}-\vec k_{\perp}|\right)\delta^2(p_{\shortparallel}-k_{\shortparallel})\,.
\end{equation}
  The decay rate of the ground state axions is then at the leading order, after multiplying the probability for the initial axion to have the radial momentum, $\vec k_{\perp}$, 
\begin{equation}
\Gamma= g_{a\gamma}^2\omega_0(k_z)\left(\frac{\Phi}{\cal N}\right)^2 |m|^3\int_{|m|}\frac{d^2k_{\perp}}{(2\pi)^2}\frac{|C_{k_z}(\vec k_{\perp})|^2}{\sqrt{ k_{\perp}^2-|m|^2}}\,,
\label{decay}.
\end{equation}
where the integration is supported only for $|\vec k_{\perp}|\ge|m|$. 

The excited states of bound axions as well will decay either directly to photons, whose rate is similar to that of the ground state, $\Gamma$, given in Eq.~(\ref{decay}), or to the low-lying states by radiating photons. 
\begin{figure}[ht]
\vskip 0.2in
\centering 
\includegraphics[scale=0.45]{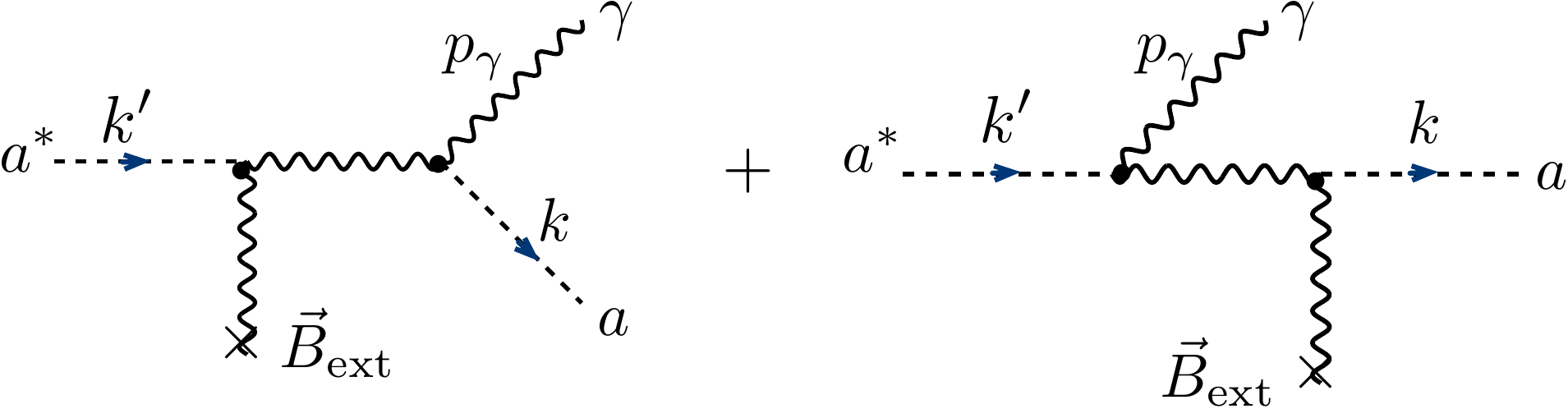}
\caption{An excited state $a^*$ decays into a lower state $a$, emitting a single photon.  }
\label{axion_decay1}
\end{figure}
The radiative transition rate to a low-lying state is given as
\begin{equation}
\Gamma_{\rm rad}= \int \frac{d^2{\vec k^{\prime}_{\perp}}}{(2\pi)^2}|C_{k_z^{\prime}}^{\prime}(\vec k_{\perp}^{\prime})|^2
\int{d\Pi_2}\,|C_{k_z}(\vec k_{\perp})|^2\left|M(a^*\to a+\gamma)\right|^2\,,
\end{equation}
where $M(a^*\to a+\gamma)$ is the invariant matrix element for the radiative transition, shown in Fig.~\ref{axion_decay1} for the leading order. $C_{k_z^{\prime}}^{\prime}(\vec k_{\perp}^{\prime})$ and $C_{k_z}(\vec k_{\perp})$ correspond to the Fourier-transform of the radial wave-functions of the excited axion $a^*$ and the lower state $a$, respectively. The phase space for the 2-particle final states
\begin{equation}
\int \!d\Pi_2=	\int\frac{d^3k}{(2\pi)^32k_0}\int\frac{d^3p_{\gamma}}{(2\pi)^32p_0}	\frac{(2\pi)^3}{|m|}\delta\left(|m|-\left|\vec k_{\perp}^{\prime}-\vec p_{\gamma\perp}-\vec k_{\perp}\right|\right)\delta^2(k_{\shortparallel}^{\prime}-p_{\gamma\shortparallel}-k_{\shortparallel})\,.
\end{equation}
One finds that the rate for the radiative transition to lower-lying states is much smaller than the decay rate of the ground state, suppressed by $(g_{a\gamma}m)^2$\,: 
\begin{equation}
\Gamma_{\rm rad}\sim g_{a\gamma}^4m^4\left(\frac{\Phi}{\cal N}\right)^2k_0^{\prime}\sim g_{a\gamma}^2m^2\Gamma\ll \Gamma\,.	
\end{equation}
The excited states therefore hardly transit to low-lying states but decay directly to photons at the rate of order of $\Gamma$\,.

\section{Applications and conclusions}

\begin{figure}[ht]
\vskip 0.2in
\centering 
\includegraphics[scale=0.45]{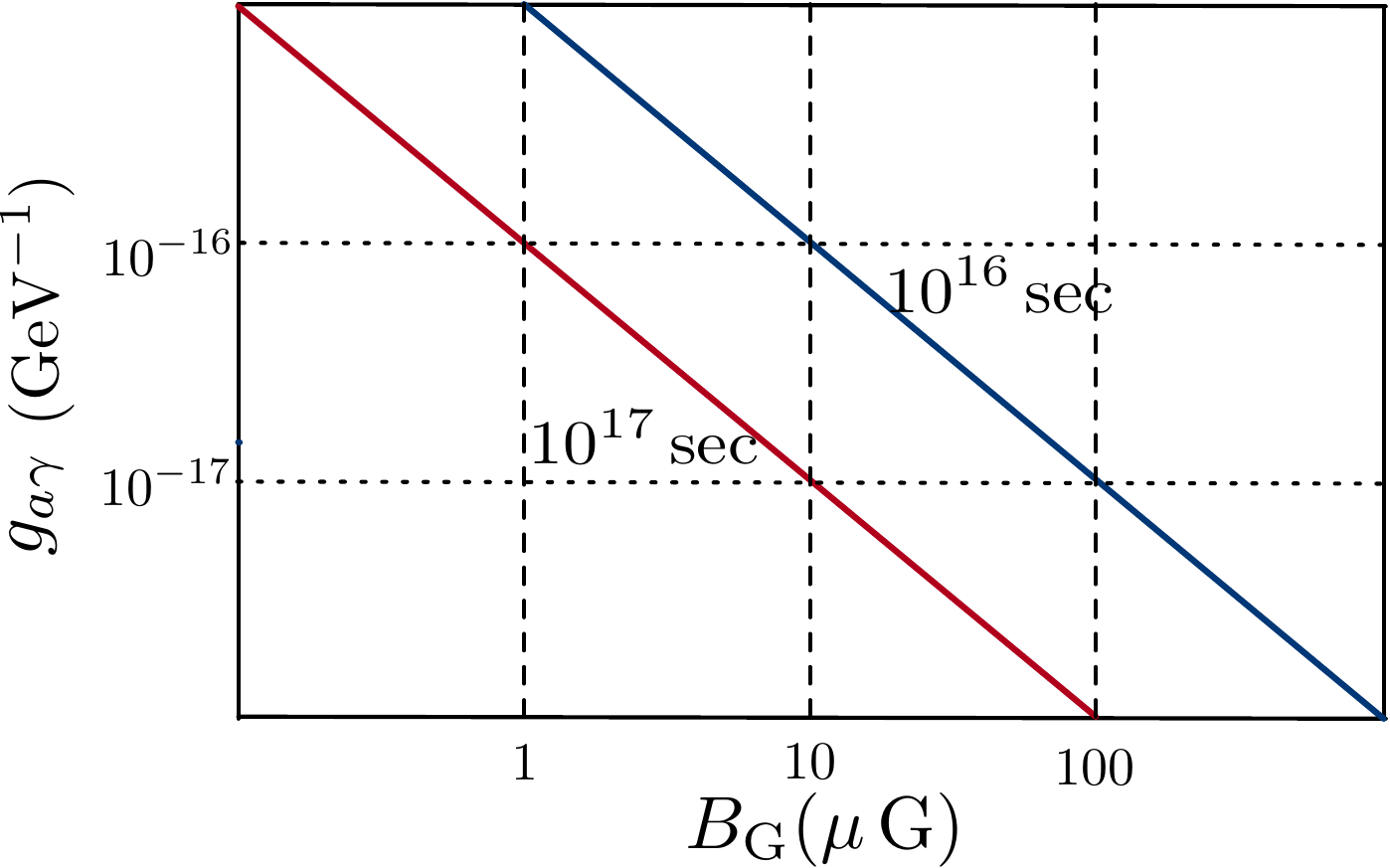}
\caption{We plot the lifetime of axions, $10^{17}\,{\rm sec}$ and $10^{16}\,{\rm sec}$,  respectively, in the magnetic axion vortex as a function of the galactic magnetic field, $B_{\rm G}$, at the center of the vortex and the axion-photon coupling, $g_{a\gamma}$. }
\label{galatic_mag}
\end{figure}

The characteristic size of the magnetic axion vortex is given by $m^{-1}=(g_{a\gamma}\dot a_0)^{-1}$, which is of galactic scale, about a few hundreds parsecs, for QCD axion dark matter, if we assume the axion kinetic energy is about the dark matter energy density in Galaxy, $\frac12\dot a_0^2\approx\rho_{\rm dm}$ :
\begin{equation}
	m^{-1}=250\,{\rm pc}\cdot\left(\frac{{10^{-14}\,\rm GeV}^{-1}}{g_{a\gamma}}\right)\cdot\left(\frac{\sqrt{0.8~{\rm GeV}{\rm cm}^{-3}}}{\dot a_0}\right)\,.
\end{equation}
Since magnetic fields are known to be ubiquitous in galaxies~\cite{Beck:2000dc}, it is quite natural to form the magnetic axion vortices in galaxies if axions are the main component of dark matter.  The magnetic fields in our Galaxy is for example estimated to have a magnitude of around a few micro Gauss with directions along the arm~\cite{Parker:1958zza}. The magnetic field at the center of the vortex is from Eq.\,(\ref{magnet}) parallel to the vortex with magnitude 
\begin{equation}
B_z(0)=m^2\frac{\Phi}{\cal N}\simeq10\,\mu\,{\rm G}\left(\frac{\Phi/{\cal N}}{10^{44}}\right)\cdot\left(\frac{m}{10^{-35}{\rm GeV}}\right)^2\,.	
\label{mag_core}
\end{equation}
For the typical parameters of QCD axion dark matter, namely for $m_a\sim 1~\mu\,{\rm eV}$, $g_{a\gamma}\simeq10^{-14}\,{\rm GeV}^{-1}$ and the axion energy density $\rho_a=0.4\,{\rm GeV}{\rm cm}^{-3}$, the magnetic flux of the vortex that has $10\mu{\rm G}$ fields at the core is about $10^{44}$.  The correction to the inertia of the normal mode is therefore quite small for typical QCD axion dark matter, 
\begin{equation}
	\delta H\equiv g_{a\gamma}^2m^2\left(\frac{\Phi}{\cal N}\right)^2\simeq10^{-10}\left(\frac{g_{a\gamma}}{10^{-14}{\rm GeV}^{-1}}\right)^2\cdot\left(\frac{m}{{10^{-35}\rm GeV}}\right)^2\cdot \left(\frac{\Phi/{\cal N}}{10^{44}}\right)^2\,.
\end{equation}
The frequency of ground state axions in the vortex is then very close to the axion mass:
\begin{equation}
\omega_0^2=m_a^2\,(1-\varepsilon^2)\,,	
\end{equation}
where $\varepsilon^2\sim\delta H$.
The radial wave function of the ground state is hence approximately proportional to the zeroth-order modified Bessel of the second kind,
\begin{equation}
R_{0k_z0}(\rho)\simeq \frac{\varepsilon m_a}{\sqrt{\pi}}K_0(\varepsilon m_a\rho)\,
\end{equation}
and its Fourier-transform
\begin{equation}
C_{k_z}(\vec k_{\perp})\simeq \frac{2\sqrt{\pi}\varepsilon m_a}{|\vec k_{\perp}|^2+\varepsilon^2 m_a^2}\,.	
\end{equation}
The decay rate of the axions\,\footnote{Axion decay considered here is similar to the axion conversion under an external magnetic field, studied in~\cite{Sikivie:1983ip}.  The effect is, however, enhanced here, because the radial wave functions of the bound axions are localized near the core of the vortex, as shown in Fig.~\ref{shape}\,(b).
} 
in the ground state is then from Eq.~(\ref{decay})
\begin{equation}
\Gamma\simeq\frac{\pi}{2}\, \omega_0	g_{a\gamma}^2 \left(\frac{\Phi}{\cal N}\right)^2\!\frac{|m|^3}{\varepsilon m_a}\approx \frac{\pi}{2}g_{a\gamma}B_z(0)=1.7\,{\rm sec^{-1}}\left(\frac{g_{a\gamma}}{{\rm GeV^{-1}}}\right)
\cdot\left(\frac{B_z(0)}{10\mu{\rm G}}\right)\,,
\label{decay_rate}
\end{equation}
where we take $\varepsilon \approx g_{a\gamma}|m|\Phi/{\cal N}$ and the ground state energy to be close to the axion mass, $\omega_0\simeq m_a$, plugging in the formula for the vortex magnetic field, Eq.~(\ref{mag_core}). 
We see that the lifetime of axions in the magnetic axion vortex depends only on the product of $g_{a\gamma}$ and $B_z(0)$. It is shorter than the age of our universe unless the axion-photon coupling is smaller than $10^{-17}{\rm GeV^{-1}}$, if the magnetic field at the core of the vortex $B_z(0)=10\,\mu {\rm G}$.  In the case of QCD axion dark matter the axion has to be therefore very light $m_a<10^{-9}{\rm eV}$.~\footnote{Being a candidate of dark matter, the QCD axion is tightly constrained by the cosmological data. For example, the initial misalignment angle has to be fine-tuned in order for very light QCD axions to explain the relic density of dark matter~\cite{Marsh:2015xka}. } 
The observed magnetic field in galaxies excludes the axion-photon coupling larger than $10^{-17}{\rm GeV^{-1}}$, provided that the magnetic fields in galaxies support magnetic axion vortices.  Otherwise, dark matter axions in galaxies decay too quickly (see. Fig.~\ref{galatic_mag})\,.

To conclude, we have shown that the axion electrodynamics admits a stable soliton solution that carries a constant magnetic flux in a medium of homogeneous axion fields with a constant time-derivative. 
The soliton, named as magnetic axion vortex, is a minimum energy configuration for a given magnetic flux that is topologically conserved. We also showed the vortex is stable under the small fluctuations of axion fields. Axions are bound in the vortex, because their energy is smaller than their mass in vacuum. Finally we have calculated the decay rate of axions, bound in the vortex, to photons. We find that the lifetime of axions in the magnetic axion vortex is much shorter than the age of our universe unless the axion-photon coupling $g_{a\gamma}<10^{-17}\,{\rm GeV^{-1}}$\, or mass $m_a<10^{-9}{\rm eV}$ for QCD axions, if we interpret the observed galactic magnetic field of a few $\mu\,G$ to be of same order of the magnetic field at the core of the vortex.


\acknowledgments
We thank D. Chung, C. Grojean, S.~H. Im, M. Neubert and D. Ryu for useful discussions. 
One of us (D.\,K.~Hong) is thankful to CERN for the hospitality during his visit. 
This research was supported by Basic Science Research Program 
through the National Research Foundation of Korea (NRF) funded by the Ministry of Education (NRF- 2017R1D1A1B06033701) and also by the Korea government (MSIT) (2021R1A4A5031460).

\end{document}